\documentclass[12pt,fleqn]{article}
\usepackage{a4wide,amsmath}
\begin{document}


\newcommand{\nl}{\notag\\}
\newcommand{\eqn}[1]{Eq.~(\ref{#1})}
\newcommand{\umu}{^{\mu}}
\newcommand{\lmu}{_{\mu}}
\newcommand{\suml}{\sum\limits}
\newcommand{\prol}{\prod\limits}
\newcommand{\inty}{\int d\mu(y)}
\newcommand{\intk}{\int_K}
\newcommand{\intl}{\int_L}
\newcommand{\intlk}{\int\limits_K}
\newcommand{\intll}{\int\limits_L}
\newcommand{\intinfi}{\int\limits_{-i\infty}^{+i\infty}}
\newcommand{\al}{\alpha}
\newcommand{\be}{\beta}
\newcommand{\ga}{\gamma}
\newcommand{\Ga}{\Gamma}
\newcommand{\ep}{\epsilon}
\newcommand{\la}{\lambda}
\newcommand{\si}{\sigma}
\newcommand{\om}{\omega}
\newcommand{\vn}{\vec{n}}
\newcommand{\vm}{\vec{m}}
\newcommand{\vk}{\vec{k}}
\newcommand{\vt}{\vartheta}
\newcommand{\Qw}{Q_{\scriptscriptstyle W}}
\newcommand{\Qfp}{Q_{\scriptscriptstyle F}^{\scriptscriptstyle\prod}}
\newcommand{\Qfs}{Q_{\scriptscriptstyle F}^{\scriptscriptstyle\sum}}
\newcommand{\brfr}[2]{\left({{#1}\over{#2}}\right)}
\newcommand{\half}{{\textstyle\frac{1}{2}}}
\newcommand{\halfpi}{{\textstyle\frac{\pi}{2}}}
\newcommand{\Df}{d\mu(f)}
\newcommand{\Tr}[1]{\textrm{Tr}\left(#1\right)}
\newcommand{\Exp}[1]{\textrm{E}\left[#1\right]}
\newcommand{\Var}[1]{\textrm{V}\left[#1\right]}


\title{{\bf Gaussian limits for discrepancies.\\
I: Asymptotic results}}

\author{
Andr\'{e} van Hameren\thanks{andrevh@sci.kun.nl}\:
and Ronald Kleiss\thanks{kleiss@sci.kun.nl}\\
University of Nijmegen, Nijmegen, the Netherlands\\
\and
Jiri Hoogland\thanks{jiri@cwi.nl}\\
CWI, Amsterdam, the Netherlands}
\maketitle

\begin{abstract}
  We consider the problem of finding, for a given quadratic measure of
  non-u\-ni\-for\-mi\-ty of a set of $N$ points (such as $L_2$
  star-discrepancy or diaphony), the asymptotic distribution of this
  discrepancy for truly random points in the limit $N\to\infty$. We
  then examine the circumstances under which this distribution
  approaches a normal distribution.  For large classes of
  non-uniformity measures, a Law of Many Modes in the spirit
  of the Central Limit Theorem
 can be derived.
\end{abstract}

\thispagestyle{empty}


\newpage
\pagestyle{plain}
\setcounter{page}{1}
\tableofcontents


\section{Introduction}
In the field of numerical integration, there are two aspects of the
general problem which bear on the accuracy of the numerical result.
The first is of course the behaviour of the integrand: typically,
wildly fluctuating functions are integrated with less accuracy than
relatively smooth ones, for the same number of integration points.
The second one is the distribution of the set of points at which one
evaluates the integrand. It stands to reason that, if one has no {\it
  a-priori\/} knowledge of the integrand, a set of points that is
fairly uniformly distributed may be expected to do better than one in
which many points cluster together. It is therefore useful to quantify
and study the notion of `uniformity of point sets', and this has been
the topic of a great number of publications \cite{prepubs,nieder1}.
The most important of such notions are those of the {\it
  star-discrepancy\/} and {\it $L_2$ star-discrepancy\/}, and more
recently other measures of non-uniformity that go under the name of
{\it diaphony\/} have been introduced as well \cite{diaphony}. In this
paper, we shall call all such measures `discrepancies'.

As has been shown in Ref.~\cite{kleiss,jhk}, the use of a particular
discrepancy in assessing the uniformity of a given point set implies
that one has some notion of the generic behaviour of the integrand: it
is tacitly assumed that the integrands to be attacked belong to some
class of functions. The particular discrepancy is then recognized as
the {\it average-case complexity\/} of the integration problem over
that function class \cite{woz,pas}.

While the Monte Carlo method, in which the integration points are
chosen at random, has long been recognized as a robust and useful way
of evaluating multivariate integrals, its relatively slow convergence
has inspired a search for other point sets whose discrepancy is lower
than that expected for truly random points. Such {\it low-discrepancy
  point sets\/} and {\it low-discrepancy sequences\/} have developed
into a veritable industry, and sequences with (asymptotically,
for large $N$) very
low discrepancy are now available, especially for problems with very
many variables \cite{lowdis}. For point sets that are extracted as the
first $N$ elements of such a sequence, though, one is usually still
compelled to compute the discrepancy numerically, and compare it to
the expectation for random points in order to show that the point set
is indeed `better than random'. This implies, however, that one has to
know, for a given discrepancy, its expectation value for truly random
points, or preferably even its probability density.  In
Refs.~\cite{jiri,hk1,hk2,hk3} we have solved this problem for large
classes of discrepancies. Although computable, the resulting
distributions are typically not very illuminating. The exception is
usually the case where the number of dimensions of the integration
problem becomes very large, in which case a normal distribution often
arises \cite{jhk,leeb}.  In this paper, we investigate this phenomenon
in more detail, and we shall describe the conditions under which this
`law of large dimensions' applies.

The layout of this paper is as follows. In section 2, we define the
general structure of a discrepancy related to a class of integrands of
which it is an average-case complexity.  We show how to derive the
probability density of this discrepancy when viewed as a stochastic
variable defined on sets of truly random points.  Then, we investigate
the conditions under which this density approaches a normal density.
Finally, in section 3 we apply our results to a few toy-models and
standard choices of discrepancy.  A number of technical points are
collected in the various Appendices.  Throughout this discussion, we
shall only consider the asymptotic limit of a very large number of
integration points. This implies that, in this paper, we cannot make
any statements on how the number of points has to approach infinity
with respect to the number of dimensions, as was for instance done in
Ref.~\cite{leeb}. In Ref.~\cite{nextpaper}, we repair this defect, and
shall be able to show which precise combination of limits has to be
taken.


\section{General definitions and statements}
To set the stage, we shall always consider the integration region to
be the $s$-dimensional unit hypercube $K=[0,1)^s$. The point set $X_N$
consists of $N$ points $x_k\umu$, where $k=1,2,\ldots,N$ labels the
points and $\mu=1,2,\ldots,s$ their co-ordinates.


\subsection{Quadratic discrepancy and complexity}
We will define quadratic discrepancies as the average-case complexity
of an integration problem in terms of its averaged squared integration
error\cite{woz}.  For the given class of real-valued functions $f(x)$,
with $x\in K$, let a measure $\Df$ on the class of functions be given,
such that the one- and two-point connected Green's functions are given
by
\begin{align}
  &\int f(x)\,\Df \;=\; 0
  \;\;,\nl
  &\int f(x_1)f(x_2)\,\Df 
  \;=\; 
  \intl h(y;x_1)h(y;x_2)\,d\mu(y)
  \;\;.
\end{align}
Here we assume, that we can define a function $h(y,x)$ and a measure
$d\mu(y)$ over some space $L$ such that the above expression makes
sense.  The variable $y$ has to be suitably defined; it may be a
continuous variable with a continuous integration measure $d\mu(y)$,
or a discrete variable, in which case $\inty$ reduces to a sum over an
enumerable set of discrete values, such as a lattice: all cases we
will consider in this article can be expressed in terms of an
enumerable set of discrete values. For the moment we will stick to the
more general notation of $d\mu(y)$.
we define the quadratic discrepancy\footnote{Note that we have taken
  a factor $N$ out of the definition of the discrepancy compared to
  other definitions in the literature. This has the
  advantage that the discrepancy
  averaged over the ensemble of truly random point sets is independent of
  $N$.}
$D_N$ as follows \cite{kleiss}:
\begin{align}
  &\frac{D_N}{N}
  \;=\;
  \int\,\eta^2\left[f\right]\,\Df
  \;\;,\\
  &\eta\,[f]
  \;=\;
  \frac{1}{N}\suml_{k=1}^Nf(x_k) - \intk\,f(x)\,dx
  \;\;.
\end{align}
In Ref.~\cite{kleiss} it was shown that, if the function measure
$\Df$ is Gaussian in the sense that the only non-vanishing connected
Green's function is the two-point function, then the integration error
will be normally distributed with zero mean and variance equal
to $D_N/N$. The discrepancy $D_N$ can be written as
\begin{align}
  &D_N 
  \;=\; 
  \frac{1}{N}\suml_{k,l=1}^N \be(x_k,x_l)
  \;\;,\nl
  &\be(x_k,x_l)
  \;=\;
  \intl \om(y;x_k)\om(y;x_l)\,d\mu(y)
  \quad,\quad
  \om(y;x_k) 
  \;=\; 
  h(y;x_k) - \intk h(y;x)\,dx
  \;.
\end{align}
In fact, $D_N$ measures how well the function $h(y;\cdot)$ is
integrated by the point set $X_N$, averaged over
$y$.  Notice that $D_N$ is nonnegative by construction, and that for
an infinite equidistributed sequence, $\lim_{N\to\infty}D_N/N=0$.
Moreover, the expected value of $D_N$ for a set of $N$ truly random
points in $K$ is given by
\begin{align}
  \Exp{D_N} 
  \;=\;& 
  \int\,\Var{f}\,\Df
  \;=\; 
  \intl\intk\om(y;x)^2\,dx\,d\mu(y)
  \;\;,
\end{align}
where $\Exp{\cdot}$ denotes the expectation value w.r.t. the uniform
distribution over the ensemble of truly random point sets with $N$
points, and $V[f]$ is the variance of the function $f(\cdot)$.
We shall always assume this expectation value to be a finite
quantity, otherwise this discrepancy cannot meaningfully be used for
truly random points.

In our approach to the calculation of discrepancy distributions, we will also
use the higher momenta $\Exp{D_N^m}$ $(m=1,2,3,\ldots)$, which
therefore have to be assumed to be finite\footnote{For the
  discrepancies we discuss, this is a valid assumption.}.  We will also
define some useful functions $\be_k$ and $\Ga_k$:
\begin{align}
  \be_k(x_1,x_2) 
  \;=\;& 
  \intk \be(x_1,x)\be_{k-1}(x,x_2)\,dx
  \;\;,\\ 
  \Ga(y_1,y_2) 
  \;=\;&
  \intk \om(y_1;x)\om(y_2;x)\,dx
  \;\;,\\
  \Ga_k(y_1,y_2) 
  \;=\;& 
  \intl \Ga(y_1,y)\Ga_{k-1}(y,y_2)\,d\mu(y)
  \;\;,
\end{align}
with $\be_1=\be$ and $\Ga_1=\Ga$. The function $\Ga$ is in a certain
sense dual to the function $\be$. It will be more convenient to use,
because the variable $y$ is often an element of a countable set and
$\Ga$ can then be viewed as a matrix, with $\Ga_k(y_1,y_2)=\Ga(y_1,y_2)^k$.


\subsection{Gaussian measures on a countable basis}
In this paper, we shall consider function classes with functions $f$
that can be written as linear combinations of a countable
set of basis functions $\{u_n\}$:
\begin{equation}
  f(x) \;=\; \sum_{n}v_nu_n(x)
  \;\;.
\end{equation}
Often we will refer to the basis functions as {\it modes\/}.  We
assume that integrals over combinations $u_{n_1}(x)u_{n_2}(x)\cdots$
exist and introduce the parameters
\begin{equation}
  w_n 
  \;=\; 
  \intk u_n(x)\,dx 
  \quad\textrm{and}\quad 
  a_{m,n} 
  \;=\; 
  \intk u_m(x)u_n(x)\,dx
  \;\;.
\end{equation}
The variance of $f$ can then be written as
\begin{align}
  \Var{f} 
  \;=\; 
  \intk f(x)^2\, dx - \left(\intk f(x)\, dx\right)^2 
  \;=\;
  \sum_{m,n} v_mv_n \left(a_{m,n}-w_mw_n\right)
  \;\;.
\end{align}
A Gaussian measure on the class of functions is obtained by taking 
\begin{equation}
  \Df 
  \;=\;
  \prod_{n}\frac{\exp(-v_n^2/2\si_n^2)}{\sqrt{2\pi\si_n^2}}\,dv_n
  \;\;.
\end{equation}
For the measure to be suitably defined, the strengths $\si_n$ have to
satisfy certain restrictions. In particular we want the functions $f$
to be quadratically integrable on the average.  The reasonable
requirement that $\Exp{D_N}$ must exist ensures that the variance of the
functions $f$ exists on the average and thus imposes a condition on
the strengths:
\begin{equation}
  \Exp{D_N}
  \;=\;
  \int \Var{f}\,\Df 
  \;=\; 
  \sum_n \si_n^2\Var{u_n} 
  \;\;.
\end{equation}
Now we can use the formalism of the previous section to construct the 
discrepancy. The two-point connected Green's function is given by 
\begin{equation}
  \int f(x_1)f(x_2)\,\Df 
  \;=\; 
  \sum_n\si_n^2u_n(x_1)u_n(x_2)
  \;\;,
\end{equation}
which is nothing but a spectral representation.
The functions $h$ and $\om$ can be taken equal to
\begin{equation}
  h_n(x) 
  \;=\; 
  \si_nu_n(x) 
  \quad,\quad
  \om_n(x) 
  \;=\; 
  \si_n(u_n(x) - w_n) 
  \;\;,
\end{equation}
where the variable $y$ is replaced by the countable index $n$. 
The function $\be$ and the matrix $\Ga$ are given by 
\begin{align}
  &\be(x_1,x_2) 
  \;=\;
  \sum_n\si_n^2(u_n(x_1)-w_n)(u_n(x_2)-w_n)
  \;\;,\\
  &\Ga_{m,n} 
  \;=\; 
  \si_m\si_n(a_{m,n} - w_mw_n)
  \;\;.
\end{align}
Note that we have for the trace of $\Ga_{m,n}$:
\begin{equation}
  \Tr{\Ga} 
  \;=\; 
  \sum\limits_{n} \si_n^2\Var{u_n} 
  \;=\; 
  \Exp{D_N}
  \;\;.
\end{equation}


\subsection{General form of discrepancy distributions}
We now turn to the problem of computing the probability density of
such a discrepancy when the $N$ points are (independently and
uniformly) randomly distributed over $K$. Introducing the Dirac
$\delta$-distribution and its representation as a Laplace transform,
we may write the probability density $H(t)$ for the value $t$ of
discrepancy $D_N=D_N(x_1,x_2,\ldots,x_N)$ as
\begin{align}
  H(t) 
  \;&=\; 
  \intk\intk\cdots\intk
  \delta\left(D_N(x_1,x_2,\ldots,x_N)-t\right)\,dx_1dx_2\cdots dx_N\nl
  \;&=\; 
  \frac{1}{2\pi i}\intinfi e^{-zt}G_0(z)\,dz
  \;\;,
  \label{formofht}
\end{align}
where the $z$ integration runs along the imaginary axis,
and $G_0(z)$ is the moment-generating function
\begin{equation}
  G_0(z) 
  \;=\; 
  \Exp{e^{zD_N}}
  \;=\; 
  \suml_{m\ge0} \frac{z^m}{m!}\,\Exp{D_N^m}
  \;\;.
\end{equation}
At this point it may be useful to note that, since $D_N$ is
nonnegative by construction, we must have $H(t)=0$ for $t<0$, and
hence no singular point of $G_0(z)$ may have a negative real part.

The task is, now, to compute $G_0(z)$ as a series expansion around
$z=0$. In Refs.~\cite{jhk,jiri,hk1,hk3} we have shown how Feynman
diagrams may be usefully employed to do this in a systematic way in
the limit of large $N$. In this paper we shall restrict ourselves to
the leading behaviour $N\to\infty$, in which limit we have
\begin{align}
  &\log(G_0(z)) 
  \;=\; 
  \suml_{k>0}\frac{(2z)^k}{2k}R_k
  \quad,\quad
  R_k 
  \;=\;
  \intk\be_k(x,x)\,dx 
  \;=\; 
  \intl\Ga_k(y,y)\,d\mu(y)
  \;\;.
\end{align}
In those cases where the $y$ variables are discrete and enumerable,
$\Ga$ can be written as a real symmetric matrix, and then we simply
have
\begin{align}
  &G_0(z) 
  \;=\;
  \left(
    \textrm{det}(1-2z\Ga)
  \right)^{-1/2}
  \quad,\quad
  R_k 
  \;=\;
  \Tr{\Ga^k}
  \;\;.
\end{align}
We shall -- symbolically -- employ the matrix and trace notation for
the continuous case as well. In general, we have
\begin{align}
  \Ga_1(y_1,y_2) 
  \;&=\; 
  A(y_1,y_2) - B(y_1)B(y_2)
  \;\;,\nl
  A(y_1,y_2) 
  \;&=\; 
  \intk h(y_1;x)h(y_2;x)\,dx
  \quad,\quad
  B(y) 
  \;=\; 
  \intk h(y;x)\,dx
  \;\;.
\label{aandb}
\end{align}
In many cases ({\it cf.\/} the case of orthonormal functions bases),
we have $B(y)=0$, but this is not necessary. In general, then,
$\Tr{\Ga^k}$ consists of $2^k$ terms. However, as shown in Appendix
A, we can combine them nicely and arrive at
\begin{align}
  G_0(z) 
  \;&=\; 
  \exp(\psi(z))/\sqrt{\chi(z)}
  \;\;,\nl
  \psi(z) 
  \;&=\; 
  \suml_{k>0}\frac{(2z)^k}{2k}\;\Tr{A^k}
  \;\;,\nl
  \chi(z) 
  \;&=\; 
  1 + \suml_{k>0}(2z)^k\;\Tr{BA^{k-1}B}
  \;\;.
\label{formofg0}
\end{align}


\subsection{Standardized variables and the Gaussian limit}
We now have derived the expression for $G_0(z)$ in the large-$N$
limit.  Given the form of $\Ga(y_1,y_2)$, we can now compute $H(t)$
for given discrepancy $t$, if only numerically; in fact this was done
for the $L_2$ star-discrepancy in Ref.~\cite{jhk} for several
dimensionalities.  In some special cases, $H(t)$ can even be given in
more-or-less closed form \cite{hk1,hk2}.  Here, however, we are
interested in possible Gaussian limits, and therefore it is useful to
replace the value $t$ of the discrepancy by the standardized variable
$\xi$, as follows:
\begin{equation}
  t \;=\; t(\xi)
  \;=\;
  \Exp{D_N} + \xi\sqrt{\Var{D_N}}
  \;\;,
\end{equation}
where the expectation $\Exp{D_N}$ and variance $\Var{D_N}$ of the
discrepancy (which equal $R_1$ and $2R_2$, respectively) are taken out
such that the stochastic variable $\xi$ always has expectation zero
and variance 1. By furthermore going over from $z$ to $u =
z/\sqrt{2R_2}$ in \eqn{formofht}, we can write the probability density
$\hat{H}(\xi)$ of $\xi$ as
\begin{align}
  \hat{H}(\xi) 
  \;&=\; 
  H(t(\xi))\frac{dt(\xi)}{d\xi}
  \nl
  \;&=\; 
  \frac{\exp(-\xi^2/2)}{2\pi i}\intinfi du\;
  \exp\left(
    \frac{1}{2}(u-\xi)^2 + 
    \suml_{k\ge3}u^k\frac{2^{(k-2)/2}}{k}\ga_k^{1/2}
  \right)
  \;\;,\nl
  \ga_k 
  \;&=\; 
  R_k^2/R_2^k
  \;\;.
  \label{hath}
\end{align}
All information on the particulars of the discrepancy are now
contained in the constants $\ga_k$, and we have that {\it the
  probability density of $\xi$ approaches the normal density whenever
  $\ga_k\to 0$ for all $k\ge3$.\/} It remains to examine under what
circumstances this can happen.


\subsection{A Law of Many Modes}
Let us assume, for the moment, that the matrix $\Ga$ is indeed a real
symmetric matrix, for instance the case of Gaussian measures
on a countable basis.  Moreover, since we know that $G_0(z)$ has no
singularities for negative values of $\textrm{Re}\,z$, the eigenvalues of
$\Ga$ are also nonnegative, and we may write
\begin{equation}
  \Tr{\Ga^k} 
  \;=\; 
  \suml_n\la_n^k
  \quad,\quad
  \ga_k 
  \;=\; 
  \left(
    \suml_n\la_n^k\right)^2\left(\suml_n\la_n^2
  \right)^{-k}
  \quad,\quad
  \la_n\ge0
  \;\;,
\end{equation}
where the various eigenvalues have been denoted by $\la_n$. Note that
the sum may run over a finite or an infinite number of eigenvalues,
but all these sums must converge since $\Exp{D_N}$ is finite. Note,
moreover, that $\ga_k$ is homogeneous of degree zero in the $\la_n$:
therefore, any scaling of the eigenvalues by a constant does not
influence the possible Gaussian limit (although it will, of course,
affect the mean and variance of $D_N$).

We now proceed by noting that $\ga_{k+1}\le\ga_k$, because
\begin{equation}
  \left( \suml_n\la_n^{k+1} \right)^2 
  \;\le\;
  \left( \suml_n\la_n^{2k} \right)
  \left( \suml_n\la_n^2 \right) 
  \;\le\;
  \left( \suml_n\la_n^k \right)^2
  \left( \suml_n\la_n^2 \right)
  \;\;,
\end{equation}
where the first inequality is simply the Schwarz inequality, and the
second one holds because the $\la_n$ are nonnegative. This means that
$\ga_k$ will approach zero for $k>3$, whenever $\ga_3$ approaches
zero. To see when this happens we define
\begin{equation}
  x_n 
  \;=\; 
  \frac{\la_n}{\sqrt{\sum_m\la_m^2}}
  \quad,\quad
  x \;=\; \max_{n} x_n
  \;\;,
\end{equation}
so that $\sum_nx_n^2=1$. It is then trivial to see that
\begin{equation}
  x^3 \;\le\; \ga_3 \;\le\; x
  \;\;,
\end{equation}
from which we derive that {\it the necessary and sufficient condition
  for the discrepancy distribution to approach a Gaussian is that}
\begin{equation}
  C 
  \;=\; 
  \frac{\la^2}{\suml_n\la_n^2} 
  \;\to\; 0
  \quad,\quad
  \la \;=\; \max_n \la_n
  \;\;.
  \label{necsuf}
\end{equation} 
The Gaussian limit is thus seen to be equivalent to the statement that
even the largest eigenvalue becomes unimportant. Clearly, a necessary
condition for this is that the total number of non-vanishing
eigenvalues (number of {\it modes\/}) approaches infinity.
Incidentally, the condition (\ref{necsuf}) also implies that
\begin{equation}
  \la \;\to\; 0
  \quad,\quad
  \suml_n\la_n^2 \;\to\; 0  
  \;\;,
\end{equation}
for all those discrepancies that have $\Exp{D_N}=\sum_n\la_n=1$.  This
is eminently reasonable, since a distribution centered around 1 and
(by construction) vanishing for negative argument can only approach a
normal distribution if its variance approaches zero.  On the other
hand, the condition $\la\to0$ is by itself {\it not\/} sufficient, as
proven by a counterexample given in Appendix B.

Another piece of insight can be obtained if we allow the eigenvalues
to take on random values. We may introduce the rather dizzying concept
of {\it an ensemble of different definitions of discrepancy,\/} each
characterized by its set of eigenvalues (all nonnegative)
$\vec{\la}=\{\la_1,\la_2,\ldots,\la_M\}$, with the usual constraint
that they add up to 1; we keep $M$ finite for simplicity.  A natural
probability measure on this ensemble is given by the probability
density $P_{\la}(\vec{\la})$ of the random vector $\la$:
\begin{equation}
  P_{\la}(\vec{\la}) 
  \;=\; 
  \Ga(M)\;
  \delta\left(\suml_{n=1}^M\la_n - 1\right)
  \;\;.
\end{equation}
Here $\Ga$ denotes Eulers gamma-function.  It is easily computed that the
expectation and variance of $R_k=\sum_n\la_n^k$ are given, for large
$M$, by
\begin{equation}
  \Exp{R_k} \sim \frac{k!}{M^{k-1}}
  \quad,\quad
  \Var{R_k} \sim \frac{(2k)!-(1+k^2)(k!)^2}{M^{2k-1}}
  \;\;,
\end{equation}
so that the $R_k$ become sharply peaked around their expectation for
large $M$. In that case, we have
\begin{equation}
  \ga_3 \;\sim\; \frac{9}{2M}
  \;\;,
\end{equation}
and we see that, in the above sense, almost all discrepancies have a
Gaussian distribution in the limit where $M$, the number of modes,
approaches infinity.



\section{Applications to different examples}
\subsection{Fastest approach to a Gaussian limit}
We now examine the various definitions of discrepancies, and assert
their approach to a Gaussian limit. Usually this is envisaged, for
instance in Ref.~\cite{leeb}, as the limit where the dimensionality $s$
of $K$ becomes very large. But, as we have shown, this is only a
special case of the more general situation where the number of
relevant modes becomes very large: another possible case is that
where, in one dimension, the number of modes with essentially equal
strength $\si_n$ becomes very large. As an illustration, consider the
case where the basis functions with the Gaussian measure are
orthonormal and $M$ of the nontrivial modes have equal strength
$\si_n^2=1/M$, and the rest have strength zero. The moment-generating
function then takes on a particularly simple form, and so does the
discrepancy distribution \cite{hk2}:
\begin{equation}
  \log(G_0(z)) 
  \;=\; 
  -\frac{M}{2}\log\left(1-\frac{2z}{M}\right)
  \quad,\quad
  H(t)
  \;=\; 
  \frac{(M/2)^{M/2}}{\Ga(M/2)}\,t^{M/2-1}e^{-tM/2}
  \;\;.
  \label{equalstrengths}
\end{equation}
It is easily seen that the gamma-distribution $H(t)$ approaches a
normal one when $M$ becomes very large. At the same time, we see the
`physical' reason behind this: it is the fact that the singularity of
$G_0(z)$ in the complex plane (in the more general case, the
singularity nearest to $z=0$) moves away to infinity.  One observation
is relevant here: in \eqn{hath}, we have kept the integration over $u$
along the imaginary axis $\textrm{Re}\,u=0$.  We might consider
performing a saddle-point integration, with a non-vanishing value of
$\textrm{Re}\,u$. That may give us, for a finite number of modes, a good
approximation to the actual form of $H(t)$.  It is quite possible,
and, indeed, it happens in the above equal-strength model, that this
approximation is already quite similar to a Gaussian.  In the
equal-strength model, a saddle-point approximation for $H(t)$ gives
precisely the form of \eqn{equalstrengths}, the only difference being
that $\Ga(M/2)$ is replaced by its Stirling approximation.  On the
other hand, for not-so-large $M$, this form is not too well
approximated by a Gaussian centered around $t=1$, since the true
maximum resides at $t=1-2/M$. Nevertheless, in this paper we are only
interested in the limiting behaviour of $H(t)$, and we shall stick to
the use of condition (\ref{necsuf}) as an indicator of the Gaussian
limit.

One interesting remaining observation is the following.  For any
finite number $M$ of eigenvalues $\la_n$ $(n=1,2,\ldots,M)$, the {\it
  smallest\/} value of the indicator $C=\la^2/\sum_n\la_n^2$ is obtained
when $\la_n=1/M$ for all $n$. In this sense, the equal-strengths model
gives, for finite $M$, that discrepancy distribution that is closest
to a Gaussian.


\subsection{$L_2$ star-discrepancy and the Wiener measure\label{wienerm}}
Here we shall discuss the standard $L_2$ star-discrepancy
\cite{nieder1}. We start with a formulation of the problem using a
continuous variable $y$ on $K$, and $d\mu(y) = dy$.  The function $h$
is given by
\begin{equation}
  h(y;x_k) 
  \;=\; 
  \prol_{\mu=1}^s\theta(x_k\umu<y\umu)
  \;\;,
\end{equation}
where we have introduced the $\theta(\cdot)$ as the logical
step-function\footnote{The logical step-function $\theta(P)$ of an
  expression $P$ is equal to 1 if the $P$ is true, and 0 if $P$ is
  false. Therefore $\theta(x<y)$ is in fact equal to the Heavyside
  function $\theta(y-x)$.}.  The Gaussian function measure
corresponding to this discrepancy is therefore seen to be defined by
\begin{equation}
  \int\,f(x_1)f(x_2)\,\Df 
  \;=\; 
  \prol_{\mu=1}^s\;\min(1-x_1\umu,1-x_2\umu)
  \;\;,
\end{equation}
which we can recognize as that variation of the standard Wiener sheet
measure in which the function $f(x)$ is pinned down at
$x=(1,1,\ldots,1)$ rather than at $x=(0,0,\ldots,0)$. This is the
content of the original Wo\'{z}niakowski lemma from Ref.~\cite{woz}.

A formulation of this discrepancy in terms of a Gaussian measure on a
countable basis can be constructed by realizing that a
spectral representation of the integration kernel
$g(x_1,x_2)=\prod_{\mu=1}^{s}\min(x_{1}^{\mu},x_{2}^{\mu})$ exists
\cite{loeve} and is given by
\begin{equation}
  \label{minspec}   
  g(x_1,x_2) 
  \;=\; 
  \suml_{\vn\ge0}\si_{\vn}^{2}\,u_{\vn}(x_1)u_{\vn}(x_2)
  \;\;,
\end{equation}
where the functions $u_{\vn}$ are given by
\begin{equation}
  \label{phi}  
  u_{\vn}(x) 
  \;=\; 
  2^{s/2}\prol_{\mu=1}^s\sin\left(
    r(n\lmu){\textstyle\frac{\pi}{2}} x\umu\right)
  \;\;,
\end{equation}
and the strengths $\si_{\vn}^{2}$ by
\begin{align}
  \label{sigmaL2}
  &\si_{\vn}^{2} 
  \;=\; 
  \left(\frac{4}{\pi^2}\right)^s
  \prol_{\mu=1}^{s}\frac{1}{r(n\lmu)^2}
  \quad,\quad
  r(n)
  \;=\; 
  (2n+1)\,\theta(n\ge0)
  \;\;.
\end{align}
Because a Gaussian measure is completely defined by its two-point Green's 
function, the measure defined by the basis functions $u_{\vn}$ is equivalent
with the Wiener measure. 
In Appendix C we show that the discrepancy defined using this formulation of 
the Gaussian measure on a countable basis is equivalent to the 
$L_2$ star-discrepancy.

The functions $u_{\vn}$ are orthonormal, and we have 
\begin{equation}
  w_n = 2^{s/2}\si_{\vn} \quad\textrm{and}\quad 
  a_{m,n} = \delta_{m,n} \;\;,
\end{equation}
where we introduced the Kronecker symbol $\delta_{m,n}$.
The matrix $\Ga$ is given by
\begin{equation}
 \Ga_{\vm,\vn} 
 \;=\; 
 \si_{\vm}^{2}\delta_{\vm,\vn}-2^s\si_{\vm}^{2}\si_{\vn}^{2}
 \;\;,
\end{equation}
and an eigenvalue equation for the eigenvalues $\la$ can be written down 
easily:
\begin{equation}
  \prod_{\vn}(\si_{\vn}^{2}-\la)
  \left[
    1-2^s\sum_{\vm}\frac{\si_{\vm}^{4}}{\si_{\vm}^{2}-\la}
  \right] 
  \;=\; 0
  \;\;.
\end{equation}
In value the strengths $\si_{\vn}$ are degenerate. Labelling the
strengths with different values by $\si_{p}$ with
$p=\prod_{\mu=1}^s r(n\lmu)$, the degeneracy is given by
\begin{equation}
  \Qw(p) 
  \;=\; 
  \suml_{\vn\geq0} 
  \theta\left(p=\prol_{\mu=1}^s r(n\lmu)\right)\;\;,
\end{equation}
so that $\la=\si_{p}^{2}$ is solution to the eigenvalue equation with
a $(\Qw(p)-1)$-fold degeneracy. If we factorize these solutions we
obtain the following equation for the remaining eigenvalues: 
\begin{equation}
  1-2^s\sum_{p}\Qw(p)\,\frac{\si_{p}^{4}}{\si_{p}^{2}-\la}
  \;=\;
  0
  \;\;.
\end{equation}
Some assertions concerning the remaining eigenvalues can be made using
this equation. On inspection, it can be seen that there are no
negative solutions, nor solutions larger than $\si_{1}^{2}$, so that
$\si_1^2$ can be used as an upper bound of the eigenvalues of $\Ga$.
If we order the $\la$ such that $\la_1\geq\la_3\geq\ldots$, then
$\si^2_1\geq\la_1\geq\si^2_3\geq\la_3\geq\ldots$. This implies that
$\Tr{\Ga^k}=\sum_{p}\Qw(p)\,\si_{p}^{2k}-\epsilon$ where
$0\leq\epsilon\leq\si_{1}^{2k}$. Note that
$\sum_{p}\Qw(p)\,\si_{p}^{2k}=\Tr{g^k}$ so that traces of
$g^k$ are upper bounds of traces of $\Ga^k$. Now we have
\begin{equation}
  \Tr{g^k} 
  \;=\; 
  \left(\frac{4}{\pi^2}\right)^{ks}\xi(2k)^s
  \quad,\quad
  \xi(p) 
  \;=\; 
  \suml_{n\ge0}{1\over(2n+1)^p}
  \;\;,  
\end{equation}
and therefore for $k\ge3$:
\begin{equation}
  \ga_k 
  \;\le\; 
  \left(\frac{\xi(2k)^2}{\xi(4)^k}\right)^s
  \left(1-2\brfr{4}{5}^s+\brfr{2}{3}^s\right)^{-k}
  \;\;.
\end{equation}
The second factor decreases monotonically from $(15)^k$ for $s=1$
to one as $s\to\infty$; for the first factor, we note that
$1<\xi(2k)<\xi(4)$ for all $k>2$. Therefore $\ga_k$ can be made
arbitrarily small by choosing $s$ large enough, and the Gaussian limit
of high dimensionality is proven. Note, however, that the approach is
not particularly fast: for large $s$, we have
$\ga_3\sim(24/25)^s\sim\exp(-s/25)$, so that $s$ has to become of the
order of one hundred or so to make the Gaussian behaviour manifest. In
fact, this was already noted by explicit numerical computation in
Ref.~\cite{jhk}.


\subsection{Diaphony}

\subsubsection{General definition}
In one dimension the discrepancy defined through a Gaussian measure on
a countable basis is called diaphony if the basis functions $\{u_n\}$
are such that
\begin{equation}
  w_n \;=\; 0 
  \quad\textrm{and}\quad 
  a_{m,n} \;=\; \delta_{m,n} 
  \;\;.
\end{equation}
These relations are typically satisfied when the functions are
orthonormal and $u_0(x)=1$ is one of the basis functions.  The matrix
$\Ga$ is given by
\begin{equation}
  \Ga_{m,n} 
  \;=\; 
  \si_{n}^{2}\delta_{m,n} 
  \;\;,
\end{equation}
so the eigenvalues are given by the squares $\si_{n}^{2}$ of the
strengths itself. An extension to more dimensions can be obtained by
taking products $u_{\vn}(x)=\prod_{\mu=1}^{s}u_{n_{\mu}}(x^{\mu})$ of
one dimensional functions. However, in contrast to the Wiener sheet
measure that underlies the $L_2$ star-discrepancy, there appears to be
no `natural' generalization of the strengths $\si_n$ to more
dimensions, and therefore we shall discuss various possibilities.  In
general, we want to let the strength $\si_{\vn}$ depend on a global
property of the vector $\vn$, for instance, the product of the
components, or the sum of the components: we shall call such
alternatives {\it clusterings\/}.


\subsubsection{Fourier diaphony}
As an application of the above, let us consider the orthonormal
functions defined by the one-dimensional factors
\begin{equation}
  u_{2k-1}(x) 
  \;=\; 
  \sqrt{2}\sin(2\pi kx)
  \quad,\quad
  u_{2k}(x) 
  \;=\; 
  \sqrt{2}\cos(2\pi kx)
  \quad,\quad 
  k=1,2,3,\ldots
  \;\;.
\end{equation}
Furthermore, it is useful to take the $\si_{\vn}$ such that the sine
and cosine modes with equal wavenumber appear with equal coefficients.
Let us define
\begin{equation}
  k(n) 
  \;=\; 
  k\,\theta\left(2k-1\le n \le 2k\right)
  \;\;.
\end{equation}
We require that $\si_{\vn}$ only depends on $\vn$ via $\vk(\vn)$:
\begin{align}
  &\si_{\vn} 
  \;=\;
  \si\left(\vk(\vn)\right)
  \quad,\quad
  \vk(\vn) 
  \;=\; 
  (k(n_1),k(n_2),\ldots,k(n_s)) 
  \;\;.
\end{align}
In that case, the diaphony is equal to
\begin{equation}
  D_N 
  \;=\; 
  \frac{1}{N}\suml_{\vk\ne0}\si^2(|k_1|,|k_2|,\ldots,|k_s|)
  \left|\suml_{l=1}^Ne^{2i\pi\vk\cdot\vec{x}_l}\right|^2
  \;\;,
\end{equation}
where, this time, the vector $\vk$ runs over the whole integer lattice
except the origin; and it has the appealing property that the value of
the Fourier discrepancy is the same for point sets differing only by a
translation mod 1; the $L_2$ star-discrepancy does not have this nice
property. 


\subsubsection{Fourier diaphony with product clustering \label{fourier}}
One of the most straightforward generalizations of the Fourier
diaphony, and the choice made in Ref.~\cite{diaphony}, is to let
$\si_{\vn}$ depend on the product of the frequency components:
\begin{align}
  &\si^2_{\vn} 
  \;=\;
  {1\over(1+\pi^2/3)^s-1}
  \prol_{\mu=1}^s\frac{1}{r(n_{\mu})^2} 
  \quad,\quad
  r(n)
  \;=\;
  \theta(n=0)+k(n)\,\theta(n>0)
  \;\;.
\end{align}
The normalization of the $\si_{\vn}$ ensures that $\Exp{D_N}=1$,
independent of $s$.  In this case, keeping in mind that sines and
cosines occur with equal strength, we have to consider the
multiplicity function
\begin{equation}
  \Qfp(p) 
  \;=\; 
  \suml_{\vn\ge0}\theta\left(p=\prol_{\mu}r(n_{\mu})\right)
  \;\;,
\end{equation}
Actually, before assigning a strength $\si_{\vn}$, or rather $\si^2_p$,
we have to know the behaviour of $\Qfp(p)$ in order to ensure
convergence of $\Exp{D_N}$. In order to do so, we introduce the
Dirichlet generating function for $\Qfp(p)$:
\begin{equation}
  F^{(1)}_s(x) 
  \;=\; 
  \suml_{p>0}\frac{\Qfp(p)}{p^x} = \left(1+2\zeta(x)\right)^s
  \;\;,
\end{equation}
where we use the Riemann $\zeta$ function. Since this function (and,
therefore, $F^{(1)}_s(x)$ as well), converges for all $x>1$, we are
ensured that $\Qfp(p)$ exceeds the value $cp^{1+\epsilon}$ at
most for a finite number of values of $p$, for all positive $c$ and
$\epsilon$. This is proven in Appendix D. It is therefore sufficient
that $\si^2_p$ decreases as a power (larger than 1) of $p$. In fact,
taking 
\begin{equation}
  \si^2_p \;=\; c p^{-\be}
  \quad,\quad
  \be \;>\; 1
  \;\;,
\end{equation}
we immediately have that
\begin{equation}
  R_k 
  \;=\; 
  \suml_{\vn>0}\si_{\vn}^{2k} 
  \;=\; 
  \suml_{p>0}\Qfp(p)\si^{2k}_p - \si^{2k}_1
  \;=\; 
  c^k\left[\left(1+2\zeta(k\be)\right)^s - 1\right]
  \;\;,
\end{equation}
which, for given $\be$, fixes $c$ such that $R_1=\Exp{D_N}= 1$, and,
moreover, gives 
\begin{equation}
  \ga_3 \sim a(\be)^s\quad\textrm{as}\quad s\to\infty
  \quad,\quad
  a(\be) 
  \;=\; 
  {\left(1+2\zeta(3\be)\right)^2\over \left(1+2\zeta(2\be)\right)^{-3}}
  \;\;.
\end{equation}
As indicated above, in Ref.~\cite{diaphony} the value $\be=2$ is used,
with $a(2)\sim 0.291$.  The supremum of $a(\be)$ equals $1/3$, as
$\be\to\infty$, and the (more interesting) infimum is $a(1)$, about
$0.147$.  We conclude that, for all diaphonies of the above type, the
Gaussian limit appears for high dimensionality. For large $\be$, where
the higher modes are greatly suppressed, the convergence is slowest,
in accordance with the observation that the `equal-strength' model
gives the fastest convergence; however, the convergence is still much
faster than for the $L_2$ star-discrepancy, and the Gaussian
approximation is already quite good for $s\sim4$. The {\it fastest\/}
approach to the Gaussian limit occurs when we force all modes to have
as equal a strength as is possible within the constraints on the
$\be$. The difference between the supremum and infimum of $a(\be)$ is,
however, not much more than a factor of $2$.

Another possibility would be to let $\si^2_p$ depend exponentially on
$p$. In that way one can ensure convergence of the $R_k$ while at the
same time enhancing as many low-frequency modes as possible.  It is
proven in Appendix D that the function
\begin{equation}
  F^{(2)}_s(x) 
  \;=\; 
  \suml_{p>0}\Qfp(p)\,x^p
\end{equation}
has radius of convergence equal to one, and therefore we may take
$\si^2_p = (\be')^p$ with $\be'$ between zero and one. If we choose
$\be'$ to be very small, we essentially keep only the modes with
$p=1$, and therefore in that case we have $\ga_3\sim1/(3^s-1)$.  This
is of course in reality the same type of discrepancy as the above one,
with $\be\to\infty$. On the other hand, taking $\be'\to1$ we arrive at
$\ga_3\to0$ (see, again, Appendix D).  The difference with the first
model is, then, that we can approach the Gaussian limit arbitrarily
fast, at the price, of course, of having a function $\be(x_k,x_l)$
that is indistinguishable from a Dirac $\delta$-distribution in
$x_k-x_l$, and hence meaningless for practical purposes.


\subsubsection{Fourier diaphony with sum clustering}
In the above, we have let the strength $\si_{\vn}$ depend on the {\it
  product\/} of the various $r(n_{\mu})$. This can be seen as mainly a
matter of expediency, since the generalization to $s>1$ is quite
simple in that case. From a more `physical' point of view, however,
this grouping of the $\si$ is not so attractive, if we keep in mind
that each $\vn$ corresponds to a mode with wave vector $\vk(\vn)$.
Under the product rule, wave vectors differing only in their direction
but with equal length may acquire vastly different weights: for
instance, $\vk = (m\sqrt{s},0,0,\ldots)$ and $\vk=(m,m,m,\ldots)$ have
equal Euclidean length, $m\sqrt{s}$, but their strengths under the
product rule are $1/(sm^2)$ and $1/(m^{2s})$, respectively. This lack
of `rotational' symmetry could be viewed as a drawback in a
discrepancy distinguished by its nice `translational' symmetry. One
may attempt to soften this problem by grouping the strengths
$\si_{\vn}$ in another way, for instance by taking
\begin{equation}
  \si_{\vn} 
  \;=\; 
  \si\left(\sum_{\mu}k(n_{\mu})\right)
  \;\;,
\end{equation}
so that $\si$ depends on the sum of the components rather than on
their product. The multiplicity of a given strength now becomes, in
fact, somewhat simpler:
\begin{equation}
  \Qfs(p) 
  \;=\; 
  \suml_{\vn>0}\theta\left(p=\suml_{\mu=1}^sk(n_{\mu})\right)
  \;=\; 
  \suml_{m\ge0}\binom{s}{m}\binom{s-1+p-m}{p-m}
  \;\;,
\end{equation}
where the last identity follows from the generating function
\begin{equation}
  F^{(3)}_s(x) 
  \;=\; 
  \suml_{p\ge0}\Qfs(p)\,x^p 
  \;=\; 
  \left(\frac{1+x}{1-x}\right)^s
  \;\;.
\end{equation}
This also immediately suggests the most natural form for the strength:
$\si^2_{\vn} = \be^p$, where $p$ is $\sum_{\mu}k(n_{\mu})$ as above.
We see that $R_1$ converges as long as $\be<1$, and moreover,
\begin{equation}
  \ga_3 
  \;=\; 
  \frac{\left[\left(\frac{1+\be^3}{1-\be^3}\right)^s-1\right]^2}
  {\left[\left(\frac{1+\be^2}{1-\be^2}\right)^s-1\right]^3}
  \;\sim\; 
  a(\be)^s
  \;\;,
\end{equation}
where $a(\be)$ has supremum $a(0)=1$, and decreases monotonically with
increasing $\be$. For $\be$ close to one, we have
$a(\be)\sim4(1-\be)/9$, so that the Gaussian limit can be reached as
quickly as desired (again with the reservations mentioned above). At
the other extreme, note that for very small $\be$ we shall have
\begin{equation}
  \ga_3 
  \;\sim\; 
  \frac{1}{2s}\quad\textrm{if}\quad s\be^2 \ll 1
  \;\;.
\end{equation}
This just reflects the fact that, for extremely small $\be$, only the
$2s$ lowest nontrivial modes contribute to the discrepancy; and even
in that case the Gaussian limit is attained, although much more
slowly.  The criterium that determines whether the behaviour of
$\ga_3$ with $s$ and $\be$ is exponential or of type $1/(2s)$ is seen
to be whether $s\be^2$ is considered to be large or small, respectively.

Another alternative might be a power-law-like behaviour of the
strengths, such as $\si^2_p = 1/p^\al$. Also in this case we may
compute the $R_k$, as follows:
\begin{equation}
  R_k 
  \;=\; 
  \suml_{p>0}\Qfs(p)\,\frac{1}{p^{k\al}} 
  \;=\;
  \frac{1}{\Ga(k\al)}\int\limits_0^{\infty}\;
  z^{k\al-1}\left(F^{(3)}_s(e^{-z})-1\right)\,dz
  \;\;,
\end{equation}
from which it follows that $\al>s$ to ensure convergence of
$\Exp{D_N}$. In the large-$s$ limit, we therefore find that, also in
this case, $\ga_3\to1/(2s)$.


\subsubsection{Fourier diaphony with spherical clustering}
A clustering choice which is, at least in principle, even more
attractive from the symmetry point of view than sum clustering, is to
let $\si_{\vn}$ depend on $|\vk(\vn)|^2$, hence assuring the maximum
possible amount of rotational invariance under the constraint of
translational invariance.  We therefore consider the choice
\begin{equation}
  \si_{\vn}^2 
  \;=\; 
  \exp\left(-\al\suml_{\mu}k(n_{\mu})^2\right)
  \;\;.
\end{equation}
For the function $\be(x_1,x_2) = \be(x_1-x_2)$ we now have the
following two alternative forms, related by Poisson summation:
\begin{align}
  \be(x) 
  & \;=\; 
  -1 + \prol_{\mu=1}^s\left(
    \suml_{k=-\infty}^{+\infty}e^{-\al k^2}\cos(2\pi
    kx^{\mu})\right)
  \nl
  & \;=\; 
  -1 + \left(\frac{\pi}{\al}\right)^{s/2}
  \suml_{\vec{m}}\exp\left(-\frac{\pi^2(\vec{x}+\vec{m})^2}{\al}\right)
  \;\;,
\end{align}
of which the first converges well for large, and the second for small,
values of $\al$; the sum over $\vec{m}$ extends over the whole integer
lattice. The $R_k$ are, similarly, given by
\begin{align}
  R_k  
  & \;=\; 
  \left(
    \suml_{q=-\infty}^{+\infty}e^{-k\al q^2}
  \right)^s-1
  \nl
  & \;=\; 
  \left(\frac{\pi}{k\al}\right)^{s/2}
  \left(
    \suml_{m=-\infty}^{+\infty}e^{-\pi^2m^2/k\al}
  \right)^s-1
  \;\;.
\end{align}
For large $\al$ (where, again, only the first few modes really
contribute) we recover, again, the limit $\ga_3\to1/(2s)$ as
$s\to\infty$: for small $\al$ we have, again, an exponential approach
to the Gaussian limit:
\begin{equation}
  \ga_3 \sim \left(\frac{8\al}{9\pi}\right)^{s/2}\quad
  \textrm{as}\quad s\to\infty
  \;\;.
\end{equation}
The distinction between the two limiting behaviours is now the
magnitude of the quantity $s\exp(-2\al)$, which now takes over the
r\^{o}le of the $s\be^2$ of the previous paragraph.


\subsubsection{Walsh diaphony}
Another type of diaphony is based on Walsh functions, which are
defined as follows. Let, in one dimension, the real number $x$ be
given by the decomposition
\begin{equation}
  x 
  \;=\; 
  2^{-1}x_1 + 2^{-2}x_2 + 2^{-3}x_3 + \cdots
  \quad,\quad
  x_i\in\{0,1\}
  \;\;,
\end{equation}
and let the nonnegative integer $n$ be given by the decomposition
\begin{equation}
  n 
  \;=\; 
  n_1 + 2n_2 + 2^2n_3 + 2^3n_4 + \cdots
  \quad,\quad
  n_i\in\{0,1\}
  \;\;.
\end{equation}
Then, the $n^{\textrm{\scriptsize th}}$ Walsh function $W_n(x)$ is defined as
\begin{equation}
  W_n(x) 
  \;=\; 
  (-1)^{(n_1x_1 + n_2x_2 + n_3x_3 + \cdots)}
  \;\;.
\end{equation}
The extension to the multidimensional case is of course
straightforward, and it is easily seen that the Walsh functions form
an orthonormal set.  The Walsh diaphony is then given by
\begin{equation}
  D_N 
  \;=\; 
  \frac{1}{N}\suml_{\vn>0}\si_{\vn}^2
  \left(\suml_{k=1}^N W_{\vn}(x_k)\right)^2
  \;\;.
\end{equation}
In Ref.~\cite{leeb}, the following choice is made:
\begin{align}
  &\si_{\vn}^2 
  \;=\; 
  \frac{1}{3^s-1}\prol_{\mu=1}^s {1\over r(n\lmu)^2}
  \;\;,\nl
  &r(n) 
  \;=\; 
  \theta(n=0) +
  \theta(n>0)
  \suml_{p\ge0} 2^{p}\,\theta\left(2^p\le n < 2^{p+1}\right)
  \;\;.
\end{align}
Note that, in contrast to the Fourier case where each mode of
frequency $n$ contains two basis functions (one sine and one cosine),
the natural requirement of `translational invariance' in this case
requires that the Walsh functions from $2^p$ up to $2^{p+1}$ get equal
strength.  The clusterings are therefore quite different from the
Fourier case.  We slightly generalize the notions of Ref.~\cite{leeb},
and write
\begin{align}
  &\si_{\vn}^2 
  \;=\; 
  \prol_{\mu=1}^s {1\over r(n_{\mu})^2}
  \;\;,\nl
  &r(n) 
  \;=\; 
  \theta(n=0) + 
  \theta(n>0)\suml_{p\ge0}
  \left(\al\be^p\right)^{-1/2}\theta(2^p\le n<2^{p+1})
  \;\;.
\end{align}
Here, we have disregarded the overall normalization of the $\si$'s
since it does not influence the Gaussian limit.  It is an easy matter
to compute the $R_k$; we find
\begin{equation}
  R_k 
  \;=\; 
  \suml_{\vn>0}\si_{\vn}^{2k} 
  \;=\;
  \left(1+\frac{\al^k}{1-2\be^k}\right)^s - 1
  \;\;,
\end{equation}
so that the requirement $\Exp{D_N} = R_1 <\infty$ implies that we must
have $\be<1/2$. Therefore, for not too small values of $\al$, we have
\begin{equation}
  \ga_3 
  \;\sim\; 
  a(\al,\be)^s 
  \quad,\quad
  a(\al,\be) 
  \;=\; 
  \frac{(1+\al^3/(1-2\be^3))^2}
  {(1+\al^2/(1-2\be^2))^3}
  \;\;.
\end{equation}
The choice made in Ref.~\cite{leeb} corresponds to $\al=1$ and
$\be=1/4$, for which we find $a(1,1/4)\sim 0.4197$. The Gaussian limit
should, therefore, be a good approximation for $s$ larger than 6 or
so.  An interesting observation is that for fixed $\be$, $a(\al,\be)$
attains a minimum at $\al =(1-2\be^3)/(1-2\be^2)$, so that the choice
$\be=1/4$ could in principle lead to $a(31/28,1/4)=0.4165$ with a
marginally faster approach to the Gaussian. The overall infimum is
seen to be $a(3/2,1/2) = 2/11 \sim 0.182$. As in the Fourier case with
product clustering and a power-law strength, there is a limit on the
speed with which the Gaussian is approached: in both cases this is
directly related to the type of clustering.

At the other extreme, for very small $\al$ we find the limiting
behaviour
\begin{equation}
  \ga_3 
  \;\sim\; 
  \frac{(1-2\be^2)^3}{(1-2\be^3)^2}\,\frac{1}{s}\quad
  \textrm{if}\quad s\al^2 \ll 1\;\;.
\end{equation}
Again in this case, the slowest possible approach to the Gaussian
limit is like $1/s$, directly related to the symmetry of the
discrepancy definition with respect to the various coordinate axes.



\subsection{Lego discrepancy}
Another class of integrands and discrepancies can be constructed by
dissecting the hypercube $K$ into $M$ non-overlapping bins $A_m$
$(m=1,2,\ldots,M)$, and taking the characteristic functions $\vt_m$ of
the bins as the basis functions of the measure. Then $w_n$ is the
volume of $A_m$, and
\begin{equation}
 \sum_{m=1}^{M}w_m \;=\; 1 
 \quad\textrm{and}\quad 
 a_{m,n} \;=\; w_n\delta_{m,n} 
 \;\;.
\end{equation}
Note that in this case $n$ runs over a finite set of values. Moreover,
this model is dimension-independent, in the sense that the only
information on the dimension of $K$ is that contained in the value of
$M$: if the dissection of $K$ into bins $A_k$ is of the hyper-cubic
type with $p$ bins along each axis, then we shall have $M=p^s$. Also,
a general area-preserving mapping of $K$ onto itself, such as the
Arnol'd cat-transform, will leave the
definition of the discrepancy invariant in the 
sense that it will
lead to a distortion (and possibly a dissection) of the various bins
$A_m$, but this influences neither $w_m$ nor (by definition) $\si_m$.
Owing to the finiteness of $M$, a finite point set can, in fact, have
zero discrepancy in this case, namely if every bin $A_m$ contains
precisely $w_mN$ points (assuming this number to be integer for every
$m$).

The matrix $\Ga_{m,n}$ has now indices that label the bins
$(m,n=1,2,\ldots M)$, where $M$ is the total number of bins:
\begin{equation}
  \Ga_{m,n} 
  \;=\; 
  \si_m\si_n\left(w_m\delta_{m,n} - w_mw_n\right)
  \;\;.
\end{equation}
We shall now examine under what circumstances the criterion
(\ref{necsuf}) for the appearance of the Gaussian limit is fulfilled.
The eigenvalues $\la_i$ of the matrix $\Ga_{m,n}$ are, of course,
given as the roots of the eigenvalue equation
\begin{equation}
  \left(
    \prol_{m=1}^M(\la_i-\si_m^2w_m)
  \right)
  \left(
    \sum_{n=1}^M\frac{w_n\la_i}{\la_i-\si_n^2w_n}
  \right) \;=\; 0
  \;\;.
\end{equation}
It is seen that there is always one zero eigenvalue (the corresponding
eigenvector has $1/\si_m$ for its $m^{\textrm{\scriptsize th}}$
component).  Furthermore the eigenvalues are bounded by
$\max_m(\si_m^2w_m)$, and this bound is an eigenvalue if there is more
than one $m$ for which the maximum is attained. At any rate, we have
for our criterion, that
\begin{equation}
  C 
  \;=\; 
  \frac{\la^2}{\suml_i\la_i^2} 
  \;\le\;
  \frac{\max_m(\si_m^2w_m)^2}{\Tr{\Ga^2}} 
  \;=\; 
  \frac{\max_m(\si_m^2w_m)^2}{\suml_{m}\si_m^4w_m^2(1-2w_m) + 
    (\suml_{m}\si_m^2w_m)^2}
  \;\;.
\end{equation}
Since the generality of the Lego discrepancy allows us to choose from
a multitude of possibilities for the $\si$'s and $w$'s, we now
concentrate on a few special cases.
\begin{enumerate}
\item {\sl All $w_m$ equal.} This models integrands whose local
  details are not resolved within areas smaller than $1/M$, but whose
  magnitude may fluctuate. In that case, we have
  \begin{equation}
    C \;<\; \frac{1}{1-2/M}\frac{(\max_m\si_m)^4}{\suml_n\si_n^4}
    \;\;,
  \end{equation}
  and a sufficient condition for the Gaussian limit is for this bound to
  approach zero. Note that here, as in the general case, only bins
  $m$ with $\si_m\ne0$ contribute to the discrepancy as well as to the
  criterion $C$, so that one has to be careful with models in which
  the integrand is fixed at zero in a large part of the integration
  region $K$: this type of model was, for instance, examined in 
  Ref.~\cite{schlier}.    
\item {\sl All $\si_m$ equal.} In this case, the underlying integrands
  have more or less bounded magnitude, but show finer detail in some
  places (with small $w$) than in other places (with larger $w$). 
  Now, it is simple to prove that 
  \begin{equation}
    C \;\le\; \frac{M\bar{w}^2}{1-2\bar{w}+1/M}
    \quad,\quad
    \bar{w} \;=\; \max_mw_m
    \;\;,
  \end{equation}
  so that a sufficient condition is that $M\bar{w}^2$ should approach zero.
\item {\sl All $\si_m^2w_m$ equal.} This choice models functions in which 
  the largest fluctuations appear over the smallest intervals. Although 
  not {\it a priori\/} attractive in many cases, this choice is actually 
  quite appropriate for, {\it e.g.\/} particle physics where cross sections 
  display precisely this kind of behaviour. In this case we simply have
  \begin{equation}
    C \;=\; \frac{1}{(M+2)(M-1)}
    \;\;,
  \end{equation}
and the Gaussian limit follows whenever $M\to\infty$.
\end{enumerate}



\section{Conclusions}
We have shown that a large class of discrepancies, including the
$L_2$ star-discrepancy and the diaphonies, can be formulated as the
{\it induced discrepancy} of a class of functions defined by a
countable set of basis functions. These basis functions we called {\it
  modes}. For such a discrepancy we derived the probability
distribution, in the limit of a large number of points, over the
ensemble of truly random point-sets.  We have shown under what
conditions this distribution tends to a Gaussian.  In particular, the
question of the limiting behaviour of a given distribution can be
reduced to solving an eigenvalue problem.  Using the knowledge
of the eigenvalues for a given function class it is possible to
determine under which conditions and how fast the Gaussian limit is
approached.  Finally, we have investigated the limiting behaviour of
the probability distribution for the discrepancy of several function
classes explicitly.

The discrepancy that most rapidly approaches the Gaussian limit occurs
for models in which the number of modes with non-zero equal strength goes
to infinity, while the sum of the strengths is fixed. In fact, we
give an argument why we cannot improve much on this limit. However, a
drawback of this model is that the discrepancy itself becomes a sum of
Dirac $\delta$-functions in this limit: it only measures whether points
in $X_N$ coincide or not, and is therefore not very useful in practice.

Secondly, we have examined the $L_2$ star-discrepancy. Here a Gaussian
distribution appears in the limit of a large number of dimensions. It
is however a very slow limit: only when the number of dimensions
becomes of the order ${\mathcal O}\left(10^2\right)$ does the Gaussian
behaviour become manifest.

For the various diaphonies, the choice of the mode-strengths is more
arbitrary. The strengths we discuss are chosen on the basis of some
preferred global properties of the diaphony, such as translation-
and/or rotation-invariance. Again for large dimensions the Gaussian
limit is attained, either as a power-law or inverse of the number of
dimension.  It is possible to choose the strengths in such a way that
the Gaussian limit is approached arbitrarily fast. But the diaphony
corresponding to that case again consists of a sum of Dirac
$\delta$-functions.

Finally, for the Lego-discrepancy, we can assign strengths to the
different modes in several ways. One possibility is to keep the product of
the squared strength and volume of the modes fixed: then, the Gaussian
limit is reached for a large number of modes.

All these results have been derived in the limit of large number of
points.  It remains to be seen however whether this is reasonable in
practice. To determine when the asymptotic regime sets in, i.e. for
which value of $N$, it is necessary to take into account the
next-to-leading contributions.  This will be the subject of
Ref.~\cite{nextpaper}.



\section*{Appendix A: The form of $G_0(z)$}
\addcontentsline{toc}{section}{Appendix A: The form of $G_0(z)$} In
this Appendix, we derive the result (\ref{formofg0}) for the form of
$G_0(z)$ in terms of the quantities $A$ and $B$ of \eqn{aandb}. For
simplicity of notation, we shall assume the discrete case where the
$A_{m,n}$ is a matrix, and the $B_{m}$ a vector; the indices $m$, $n$
are then what we called the variables $y$ in the foregoing. Moreover,
let us denote by $[BA^kB]$ the sum $\sum_{m,n}B_m(A^k)_{m,n}B_n$.
Since the matrix $\Ga_{m,n}$ can be written as
\begin{equation}
  \Ga_{m,n} \;=\; A_{m,n} - B_mB_n
  \;\;,
\end{equation}
the $k^{\textrm{\scriptsize th}}$ power of this matrix has the general
form
\begin{equation}
  (\Ga^k)_{m,n} 
  \;=\; 
  (A^k)_{m,n} -
  \suml_{p,q,\nu_{0,1,2,\ldots}\ge0}
  \frac{(\sum_{r\ge0}\nu_r)!}{\nu_0!\nu_1!\nu_2!\cdots}
  (A^pB)_m(BA^q)_n\prol_{r\ge0}(-[BA^rB])^{\nu_r}
  \;\;,
\end{equation}
with the constraint $k-1=p+q+\nu_0+2\nu_1+3\nu_2+\cdots$. The
combinatorial factor follows directly from the possible positionings
of the dyadic factors $-B_mB_n$. Multiplying by $(2t)^{k-1}$ and
summing over the $k$ then gives us immediately
\begin{equation}
  \hspace{-0.7cm}
  \Tr{\frac{\Ga}{1-2t\Ga}} 
  \;=\;
  \suml_{k\ge1}(2t)^{k-1}\Tr{A^k} +
  \frac{-1}{1+\suml_{n\ge1}(2t)^n[BA^{n-1}B]}
  \suml_{r\ge0}(r+1)(2t)^r[BA^rB]
  \;\;,
\end{equation}
where the last factor, with $r+1$, comes from the double sum over $p$
and $q$ with $p+q=r$. Upon integration of this result over $t$ from 0
to $z$ we find
\begin{align}
  \log(G_0(z)) 
  \;&=\; 
  \suml_{n>0}\frac{(2z)^n}{2n}\Tr{\Ga^n}\nl
  \;&=\; 
  \suml_{n>0}\frac{(2z)^n}{2n}\Tr{A^n} -
  \frac{1}{2}\log\left(1+\suml_{n>0}(2z)^n[BA^{n-1}B]\right)
  \;\;.
\end{align}
This result has, in fact, already been obtained for the case of the
$L_2$ star-dis\-crep\-an\-cy in Ref.~\cite{jhk}, but here we
demonstrate its general validity for more general discrepancy
measures. In those cases where $B_m=0$, the second term of course
vanishes.


\section*{Appendix B: A counterexample}
\addcontentsline{toc}{section}{Appendix B: A counterexample} In this
Appendix we prove that the condition (\ref{necsuf}) for the occurrence
of a Gaussian limit is, in a sense, the best possible.  Namely,
consider a set of eigenvalues $\la_n$, again adding up to unity as
usual, defined as follows:
\begin{align}
  \la_1 \;&=\; \la
  \;\;,&\nl
  \la_n \;&=\; (1-\la)/(M-1) \;\;,& n=2,3,\ldots,M
  \;\;,\nl
  \la_n \;&=\; 0 \;\;,& n>M
  \;\;.
\end{align}
Clearly, $\la$ will indeed be the maximal eigenvalue as long as
$M>1/\la$.  Now,
\begin{equation}
  \frac{\la^2}{\sum_n\la_n^2} 
  \;=\; 
  \frac{\la^2}{\la^2 + (1-\la)^2/(M-1)}
  \;\;,
\end{equation}
and this ratio can be driven as close to unity as desired by choosing
$M$ sufficiently large. This shows that the simple condition $\la\to0$
is not always enough to ensure the Gaussian limit.


\section*{Appendix C: Spectral representation of the $L_2$ star-discrepancy} 
\addcontentsline{toc}{section}{Appendix C: Spectral representation of
  the $L_2$ star-discrepancy} 

Mercer's theorem\cite{loeve} states that a nonnegative-definite and
continuous function on $(1,0]\times(1,0]$ has a spectral
decomposition. Applying this to the function $\min(x_1^{\mu},x_2^{\mu})$,
then tells us that the two-point connected Green's function $g$ of the
Wiener measure has a spectral decomposition of \eqn{minspec}. 
The eigenvalues $\si^2_{\vn}$ and eigenfunctions $u_{\vn}$ for $g$ are
given by \eqn{sigmaL2} and \eqn{phi}.

To show that the discrepancy defined through the functions $u_{\vn}$ is the 
same as the $L_2$ star-discrepancy pinned down at $(1,1,\ldots,1)$, we
prove the equality of the $\beta$-functions for the two measures:
\begin{align}
  \hspace*{-0.5cm}
  \sum_{\vn}
  &\si_{\vn}^{2}(u_{\vn}(x_1)-2^{s/2}\si_{\vn})
                (u_{\vn}(x_2)-2^{s/2}\si_{\vn}) 
                \nl
 &\;=\; 
 \int_{K^{\prime}}
 \left(
   \prod_{\mu=1}^{s}\theta(x_{1}^{\mu}-y^{\mu}) -
   \prod_{\mu=1}^{s}(1-y^{\mu})
 \right) 
 \left(\prod_{\mu=1}^{s}\theta(x_{2}^{\mu}-y^{\mu}) -
   \prod_{\mu=1}^{s}(1-y^{\mu})
 \right)dy 
 \;\;.
\end{align}
Evaluating both sides of the equation we obtain:
\begin{align}
  \sum_{\vn}
  \left(
    \si_{\vn}^{2}\,u_{\vn}(x_1)
  \right.
  &\left.
    u_{\vn}(x_2) \,-\, 
    2^{s/2}\si_{\vn}^{3}\,u_{\vn}(x_1) \,-\, 
    2^{s/2}\si_{\vn}^{3}\,u_{\vn}(x_2) \,+\,
    2^{s}\si_{\vn}^{4}
  \right) 
  \nl 
  &=\;
  g(x_1,x_2) - 
  \prod_{\mu=1}^{s}\left(x_1^{\mu}-\half(x_1^{\mu})^2\right) - 
  \prod_{\mu=1}^{s}\left(x_2^{\mu}-\half(x_2^{\mu})^2\right) +
  \left(\frac{1}{3}\right)^s 
  \;\;.
\end{align}
The first terms on both sides of the equation cancel trivially.  
A small calculation shows that the same applies to the last terms on
both sides of the equation.
It thus remains to show that 
\begin{equation}
\sum_{\vn}2^{s/2}\si_{\vn}^{3}\,u_{\vn}(x) \;=\;  
\prod_{\mu=1}^{s}(x^{\mu}-\half (x^{\mu})^2) \;\;.
\label{Fouriercoeff}
\end{equation}
This problem again factorizes for the different coordinates 
(omitting indices):
\begin{equation}
  2\left( \frac{2}{\pi} \right)^3
  \sum_{n=0}^{\infty}  \frac{1}{(2n+1)^3}\,
  \sin\left((2n+1)\halfpi x\right)  
  \;=\; 
  x-\half x^2 
  \;\;,
\label{fourdec}
\end{equation}
which is nothing but stating that the lhs of \eqn{fourdec} is
the Fourier decomposition of the rhs. To prove this, let $f$ be the
following periodic extension of $x-\half x^2$\,:
\begin{equation}
  f(x) \;=\; 
  \begin{cases}
      \half x^2-3x+4 
      &x\in(2m-2,2m]
      \\
      x-\half x^2  
      &x\in(2m,2m+2]
  \end{cases}    
\end{equation}
where $m$ is any integer.
The function $f$ is a parabolic approximation of $\half\sin(4\pi x)$. 
It is continuous and differentiable on $\mathbf{R}$. Hence it can be written 
as a Fourier series, based on a period of $4$ rather than $1$. An explicit 
calculation shows that the only non-zero terms comes from the functions 
$\frac{1}{\sqrt{2}}\sin\left((2n+1)\halfpi x\right)$  $(n=0,1,2,\ldots)$.
The Fourier coefficients are given by 
\footnote{We take the functions normalized such that they form a orthonormal 
set on $(0,4]$, so the Fourier series is in terms of the sine- and cosine 
functions divided by $\sqrt{2}$.}
\begin{equation}
  \frac{1}{\sqrt{2}}\int_0^{4} f(x)\sin\left((2n+1)\halfpi x\right)\,dx 
  \;=\;
  2\sqrt{2}\left(\frac{2}{\pi}\right)^3\frac{1}{(2n+1)^3} 
  \;\;.
\end{equation}
Thus the Fourier series is exactly given by the lhs of \eqn{fourdec}.


\section*{Appendix D: The magnitude of $\Qfp(p)$}
\addcontentsline{toc}{section}{Appendix D: The magnitude of
  $\Qfp(p)$} Here we present the proofs of our various
statements about the multiplicity function $\Qfp(p)$ of section
\ref{fourier}.  In the first place, we know that its Dirichlet
generating function, $F^{(1)}(x)$, converges for all $x>1$. Now
suppose that $\Qfp(p)$ exceeded $cp^{\al}$ an infinite number of
times, with $c>0$ and $\al>1$. The Dirichlet generating function would
then contain an infinite number of terms all larger than $c$, for
$1<x<\al$, and therefore would diverge, in contradiction with its
convergence for all $x>1$.

In the second place, consider the `standard' generating function,
$F^{(2)}_s(x)$. By inspecting how many of the vector components
$n_{\mu}$ of $\vn$ are zero, we see that we may write, for $p>1$,
\begin{equation}
  \Qfp(p) 
  \;=\; 
  \suml_{t=1}^s\binom{s}{t}2^td_t(p)
  \quad,\quad
  d_t(p) 
  \;=\; 
  \suml_{\vn\ge0}\theta\left(p=\prol_{\mu=1}^tn_{\mu}\right)
  \;\;,
\end{equation}
so that $d_t(p)$ counts in how many ways the integer $p$ can be
written as a product of $t$ factors, including ones; this function is
discussed, for instance, in Ref.~\cite{hardy}. Now, for $p$ prime, we
have $d_t(p)=t$, and therefore
\begin{equation}
  \Qfp(p) 
  \;\ge\; 
  2s(3^{s-1})
  \quad,\quad\textrm{equality for $p$ prime}
  \;\;.
\end{equation}
The radius of convergence of $F^{(2)}_s(x)$ is therefore {\it at
  most\/} equal to unity.  On the other hand, we can obtain a very
crude, but sufficient, upper bound on $\Qfp(p)$ as follows.
Since $d_t(p)$ is a nondecreasing function of $t$, we may bound
$\Qfp(p)$ by $(3^s-1)d_s(p)$. Now let $k_p$ be the number of
prime factors in $p$; then $k_p$ cannot exceed $\log(p)/\log(2)$, and
only is equal to this when $p$ is a pure power of 2. Also, the number
of ways to distribute $k$ object in $s$ groups (which may be empty) is
at most $s^k$, and is smaller if some of the objects are equal.
Therefore, $d_s(p)$ is at most $s^{k_p}$, and we see that
\begin{equation}
  \Qfp(p) 
  \;<\; 
  (3^s-1)p^{\log(s)/\log(2)}
  \;\;,
\end{equation}
or, in short, is bounded\footnote{Note that equality cannot occur in
  this case since the two requirements are mutually exclusive.}  by a
polynomial in $p$.  Therefore, the radius of convergence of
$F^{(2)}_s(x)$ is also {\it at least\/} unity, and we have proven the
assertion in Eq.~\ref{fourier}.

Finally, we consider the limit
\begin{equation}
  \lim_{\be'\to1}\ga_3 
  \;=\; 
  \lim_{x\to1}
  \frac{\left(F^{(2)}_s(x^3)\right)^2}{\left(F^{(2)}_s(x^2)\right)^3}
  \;\;.
\end{equation}
The same reasoning that led us to the radius of convergence shows
that, for $x$ approaching 1 from below, the function $F^{(2)}_s(x)$
behaves as $(1-x)^{-c}$, with $c\ge1$. Therefore, $\ga_3$ will behave
as $(8(1-x)/9)^c$, and approach zero as $x\to1$.  Note that the upper
bound on $\Qfp(p)$ is extremely loose: but it is enough.




\end{document}